\documentclass[prl,twocolumn]{revtex4}
\usepackage{pstricks,pstcol}
\usepackage{amsmath,amssymb}
\usepackage{graphicx}
\usepackage[dvips,colorlinks=true,hyperfootnotes=true,urlcolor=black,linkcolor=blue,citecolor=blue]{hyperref}
\begin{document}
\title{Symmetry breaking by the sea of Dirac-Landau levels in graphene}
\date{\today}
\author{Vinu Lukose}
\email{vinu@imsc.res.in}
\author{R. Shankar}
\email{shankar@imsc.res.in}
\affiliation{The Institute of Mathematical Sciences, Chennai}

\begin{abstract} 

The quantum Hall states of graphene have a filled sea of lower Dirac-Landau
levels.  The short ranged $SU(4)$ symmetry breaking interactions can induce a
staggered polarization of the sea of Dirac-Landau levels. We study this effect
in the extended Hubbard model on a honeycomb lattice using mean field
variational wavefunctions. We find a valley symmetry broken,
anti-ferromagnetic spin ordered phase at $\nu=\pm 1$ when the on-site
interaction is dominant. Our mean field solution is consistent with the
recently reported experimental results of Z. Jiang et. al.\cite{jiang} 

\end{abstract}

\maketitle

The low energy physics of graphene is described by quasiparticles satisfying
the massless Dirac equation \cite{geim} \cite{kim}. There are two species of
Dirac fermions per spin making a total of four species of quasiparticles. In
the non-interacting limit, these four species are degenerate resulting in an
$SU(4)$ symmetry of the non-interacting theory. The dominant interaction is
the long-range part of the Coulomb interaction which is also $SU(4)$
symmetric. 

The quantum Hall effect has been observed in graphene \cite{jiang} \cite{geim}
\cite{kim} \cite{abanin} \cite{zhang} . At low magnetic fields
($\sim1$-$10~T$), plateaus are observed at $(h/e^2)\sigma_H=4(n+1/2)$
consistent with there being four degenerate Dirac-Landau levels. At higher
fields ($\sim 20$-$45~T$), this degeneracy is lifted \cite{zhang}. It is
completely lifted in the $n=0$ level resulting in extra plateaus at
$(h/e^2)\sigma_H= -1,0,1$ and partially lifted for $n=\pm1$ yielding extra
plateaus at $(h/e^2)\sigma_H=\pm 4$.  Tilted field experiments \cite{zhang}
indicate that the plateau at $(h/e^2)\sigma_H=\pm4$ is due to the Zeeman
splitting of the Landau levels. Recent experiments by Z. Jiang et. al.
\cite{jiang} show that the transport gap at the
$(h/e^2)\sigma_H=\pm1~(\nu=\pm1)$ plateau does not depend on the parallel
component of the magnetic field and is proportional to $\sqrt{B_\perp}$. These
experiments hence indicate that the $\nu=\pm1$ plateau corresponds to a state
with zero net spin polarisation. These plateaus are attributed to the lifting
of the sub-lattice degeneracy of the $n=0$ Landau level. 

It has been shown that the long range part of the Coulomb interaction causes
the $SU(4)$ symmetry to be broken spontaneously \cite{nomura} \cite{kun-yang}
\cite{alicea} \cite{goerbig}. The resulting gap is proportional to
$\sqrt{B_\perp}$. The exact pattern of the symmetry breaking is not determined
by the $SU(4)$ symmetric long range part of Coulomb interaction but by other
symmetry breaking interactions \cite{yang-review}. These could be the Zeeman
term, short range lattice scale interactions \cite{herbut} \cite{alicea}
\cite{gusynin} and disorder \cite{abanin2}.  The effects of the short range
lattice scale interactions can be studied in the extended Hubbard model on the
honeycomb lattice in the presence of a magnetic field. The previous studies of
this model \cite{alicea}\cite{herbut} find a spin polarised state at
$\nu=\pm1$. However, this is not consistent with the experiments discussed
above \cite{jiang}.

In this work, we are mainly concerned with the physics of $\nu=\pm1$ states
and we show that the sea of Dirac-Landau levels plays a significant role. We
write down mean field trial wavefunctions that incorporate a staggered $SU(4)$
polarization of the $n\neq 0$ Landau levels and compute their energies
analytically in a systematic $a/l_c$ expansion \cite{goerbig}, where
$a~(=2.45~A^\circ)$ is the lattice spacing and $l_c~(\approx 40~A^\circ~{\rm
at}~45~T)$ is the magnetic length. We then find that when the on-site
repulsion is dominant, the ground state has anti-ferromagnetic ordering with
broken valley symmetry. It has zero net spin polarisation and hence is a strong
candidate for the state found in the experiments \cite{jiang}.

The hamiltonian we consider is,

\begin{eqnarray} \label{ham-lat}
\lefteqn{ \mathcal{H}=-t\sum_{<ij>,\sigma}
\left(e^{i\phi_{ij}}c^\dagger_{i\sigma}c_{j\sigma}+h.c\right)}\\ 
\nonumber
&&+\frac{U}{2}\sum_i (\hat{n}_i-1)^2  
 + V\sum_{<ij>}(\hat{n}_i-1)(\hat{n}_j-1)
\end{eqnarray}
$\hat{n}_i=\sum_{\sigma}c^\dagger_{i\sigma}c_{i\sigma}$ is the number operator
and $\phi_{ij}$ is the phase due to presence of the magnetic field. The
continuum model describing the low energy physics can be derived in the
standard way. The continuum effective hamiltonian is,

\begin{widetext}
\begin{equation}
\label{ham-cont}
\mathcal{H} = 
\int\! d^2x ~~ \Psi^\dag  (v_F \textrm{\boldmath{$\alpha$}}
.\mathbf{\Pi}) \Psi 
 - \frac{Ua^2}{12} \left( (\Psi^\dag \sigma^a \Psi)^2 +
(\bar{\Psi} \tau^z \sigma^a \Psi)^2 + \frac{1}{2}(\Psi^\dag \alpha^i
\tau^j \sigma^a \Psi)^2 \right) 
+  \frac{3Va^2}{4} \Big( (\Psi^\dag \Psi)^2 - ( \bar\Psi 
\tau^z \Psi )^2 \Big) 
\end{equation}
\end{widetext}

Where $\mathbf{\Pi}$ is the covariant derivative.  The kinetic energy term has
an $SU(4)$ internal symmetry whereas the interaction terms break it down to
$(Z_2 \rtimes U(1))_{valley}\otimes SU(2)_{spin}$. The remanent of the $SU(2)$
valley symmetry of the non-interacting theory corresponds to 

\begin{eqnarray}
\label{z2}
\Psi_{r\eta\sigma}(\mathbf{x}) & \rightarrow &
\tau^x_{\eta\tilde{\eta}} \Psi_{r\tilde{\eta}\sigma}(\mathbf{x}) \\
\label{u1}
\Psi_{r\eta\sigma}(\mathbf{x}) & \rightarrow &
\left(e^{i\theta\tau^z}\right)_{\eta\tilde{\eta}} 
\Psi_{r\tilde{\eta}\sigma}(\mathbf{x}) 
\end{eqnarray}

where, $r$ is the Dirac (sub-lattice) index, $\eta$ the valley index and
$\sigma$ the spin index.  Note that the $Z_2$ does not commute with the
$U(1)$. The $Z_2$ part of the valley symmetry corresponds to the interchanging
of the valley indices. The $U(1)$ part corresponds to the conservation of the
number of quasiparticles belonging to the two valleys separately. This is a
symmetry only of the low energy sector as scattering from one valley to the
other is a high momentum transfer process. There are also no Umklapp processes
which contribute to this process.

Our variational wavefunctions generalise those written down by  Yang et. al.
\cite{kun-yang} to include the polarisation of the Dirac sea.  They are
constructed from the eigenfunctions of the following massive single particle
Dirac equation,

\begin{eqnarray}
\label{1pham}
h &=& v_F \, \textrm{\boldmath{$\alpha$}}.\mathbf{\Pi} + \beta mQ 
\end{eqnarray}
where $m$ is a real number and $Q$ a matrix constructed from a basis
of four orthonormal $SU(4)$ spinors, 
\begin{equation}
\label{qdef}
Q = \sum_{p=1}^k\chi^p (\chi^p)^{\dag}  
- \sum_{p=k+1}^4 \chi^p (\chi^p)^{\dag}
\end{equation}
for the case where $k$ of the four $n=0$ levels are occupied. 

The variational states are,
\begin{eqnarray}
\label{varstate}
\vert k\rangle=\left(\prod_{p=1}^k\prod_l\psi^\dagger_{0lp}\right)
\left(\prod_{n=-N_c}^{-1}\prod_{lp}\psi^\dagger_{nlp}\right)\vert 0\rangle
\end{eqnarray}
where,
\begin{equation}
\Psi(\mathbf{x})=\sum_{nlp}\Phi^{nl}(\mathbf{x})\chi^p\psi_{nlp}
\end{equation}

\begin{figure}[ht] \centering
\includegraphics[width=0.45\textwidth]{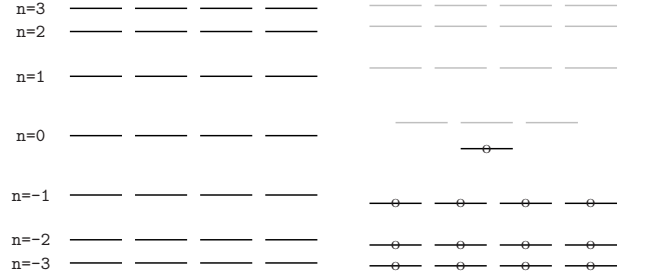}
\caption{(a) shows the energy levels for non-interacting Dirac particles in
magnetic field for graphene.(b) shows the shift in one particle levels due to
mass and the filling of one particle states for $\nu=-1$.}
\label{spectrum}
\end{figure}

$\Phi^{nl}(\mathbf{x})\chi^p$ being the eight component spinor eigenfunctions
of the hamiltonian in Eq.(\ref{1pham}). The cutoff on the number of Landau
levels, $N_c$, is obtained by matching the total number of states with that of
the lattice.  We have, $N_c= (2\pi/ \sqrt{3}) (l_c^2/a^2)$. Note that in the
limit $m\rightarrow 0$, our variational states reduce to those of Yang et.
al. \cite{kun-yang}. The ground state manifold is $U(4)/(U(k)\times U(4-k)$
for all $m$. The single particle energy levels are shifted by the presence of
the mass term as shown in Fig.(\ref{spectrum}). 

For the computation of the expectation value of the hamiltonian, we need
to compute the coincident two point correlation function 
\[\Gamma =  
\sum_{nlp\in occ} \langle
(\Phi^{nlp}(\mathbf{x}))^\dag \Phi^{nlp}(\mathbf{x}) \rangle 
\equiv\sum_{n=-N_C}^0 \Gamma^n
\]

Assuming that $m\sim (a/l_c)^2$ we put $m=\tilde m (a^2/(2\pi l_c^2))$. This
assumption will be justified later when we solve for $m$. $\Gamma^n$ can be
computed to leading order in $a/l_c$ to be,

\begin{eqnarray}
\label{gamman0}
\Gamma^0&=&\frac{1}{2\pi l_c^2}\frac{1-\beta}{2}\frac{1+Q}{2}\\
\label{gammann}
\Gamma^n&=&\frac{1}{2\pi l_c^2}\left(\frac{1}{2}+
\frac{\beta}{2}\frac{a}{2\pi l_c}{\rm sgn}(n)\frac{\frac{\tilde m}{\sqrt{3} t} Q}
{\sqrt{2 \vert n\vert}} \right)~,n\neq 0
\end{eqnarray}

Equation (\ref{gammann}) shows that the mass term induces a staggered $SU(4)$
polarization of the $n\neq 0$ Landau levels. Since the Dirac index is the same
as the sub-lattice index, the order parameters for staggered $SU(4)$ order are
\ $\langle\bar\Psi\lambda^a\Psi\rangle={\rm Tr}~\beta\lambda^a\Gamma$, where
$\lambda^a$ are the $SU(4)$ generators.

Using the two point coincident correlation function we can compute the
expectation values of hamiltonian in terms of $Q$ the matrix. The kinetic
energy density, $E_t$, interaction energy densities $E_V$ and $E_U$ are
computed to be, (here we have retained the terms upto leading order in $a/l_c$
and dropped the terms that are independent of $m$ and $Q$) 

\begin{eqnarray}
\label{et}
E_t&=&  \frac{1}{2\pi l_c^2} \frac{a^2}{2\pi l_c^2} \frac{2\tilde{m}^2}{\tilde{t}} 
\end{eqnarray}

\begin{eqnarray}
\label{ev}
E_V&=& \frac{1}{2\pi l_c^2}  \frac{a^2}{2 \pi l_c^2} 
\frac{3V}{4} \left( -\frac{\mu^2}{2} -
      \frac{\mu^2}{4} ({\rm Tr}(\tau^z Q))^2 \right.  \nonumber \\
      && \hspace{2.5cm}\left. { } + 
\frac{\mu^2+1}{8} {\rm Tr}(\tau^z Q \tau^z Q)\right)
\end{eqnarray}

\begin{eqnarray}
\label{eu}
E_U&=& \frac{1}{2\pi l_c^2}  \frac{a^2}{2 \pi l_c^2}  \frac{U}{12}
    \bigg( \frac{1}{4} ({\rm Tr}(\sigma^a Q))^2  \nonumber \\
    & & { }- \frac{\mu^2+1}{8} {\rm Tr}(\sigma^a Q \sigma^a Q) 
    + \frac{\mu^2}{4}({\rm Tr}(\tau^z \sigma^a Q))^2 \nonumber \\
    && \qquad {} -\frac{\mu^2+1}{8} 
{\rm Tr}(\tau^z \sigma^a Q \tau^z \sigma^a Q)
    \nonumber\\
    & & \qquad {}+ \frac{\mu^2-1}{8} {\rm Tr}(\tau^j \sigma^a Q \tau^j
    \sigma^a Q)\bigg)
\end{eqnarray}

where, $\mu \equiv 1 + (2\tilde{m}/\tilde{t})$ and $\tilde{t} \equiv
(t/2)(3\sqrt{3}\pi)^{1/2} $.  Note that the $m\rightarrow 0$ limit is obtained
by putting $\mu=1$ i.e. the contribution of filled sea of Dirac-Landau
levels is neglected to the mean field energy.

We now specialise to the case when $\nu=-1~(k=1)$ where a single $n=0$
Dirac-Landau level is filled. The  ground state manifold is parametrised by a 
single $SU(4)$ spinor which specifies the $SU(4)$ polarization.  The
$\nu=1~(k=3)$ case is related to the former by a particle-hole
transformation.We use an explicit parameterisation,

\begin{eqnarray}
\label{chipar}
\vert\chi\rangle &=& \cos \frac{\gamma}{2} |+\rangle |\hat n_1\rangle 
+ e^{i\Omega}\sin \frac{\gamma}{2} |-\rangle |-\hat n_2\rangle
\end{eqnarray}

This corresponds to a linear superposition of an electron with valley index
$+$, spin polarization $\hat n_1$ and valley index $-$, spin polarization
$-\hat n_2$. $\gamma$ and $\Omega$ specify the relative amplitude and
phase of the superposition. 

When the $\gamma$ is either $0$ or $\pi$ the many body state has a definite
number of electrons in each valley. The $n=0$ level electrons are localised in
one sub-lattice with arbitrary spin, corresponding  to charge and spin
ordering. These are the states discussed in previous work
\cite{alicea}\cite{herbut}. They have the $U(1)$ part of the valley symmetry
unbroken and the $Z_2$ part broken. 

When $0<\gamma<\pi$, the many body state does not have a definite number of
electrons in each valley. The total number however, remains a good quantum
number.  Further, if $\gamma \ne \pm \pi/2$, then the average number of
electrons in each valley is not the same and there is charge ordering. In this
case the full valley symmetry, $Z_2\rtimes U(1)$, is broken. When
$\gamma= \pm \pi/2$, the average number of electrons in the two valleys are
equal, the $Z_2$ symmetry is unbroken and the $U(1)$ is broken.
 
In general the state will have a non-zero total spin polarization except when
$\gamma=\pm \pi/2$ and $\hat n_1=\hat n_2$. In this case there is
anti-ferromagnetic spin order and no charge order.

We now evaluate the energy density in terms of our parameters by substituting
Eq.(\ref{chipar}) in equations (\ref{qdef},\ref{ev} and \ref{eu}).  The
total mean field energy density ($E$) is the sum of kinetic ($E_t$), nearest
neighbour interaction ($E_V$), Hubbard interaction ($E_U$) and Zeeman term
($E_Z$).

\begin{eqnarray}
\label{etot}
E &=& \frac{1}{2\pi l_c^2} \frac{a^2}{2 \pi l_c^2} 
\Bigg( \frac{2\tilde{m}^2}{\tilde{t}} 
 -  \frac{3}{4}V
\left( \frac{1}{2} + \frac{\mu^2-1}{2} (1+\cos^2 \gamma) \right)  \nonumber \\
& & \hspace{2cm} {} -  \frac{1}{16}U \left((\mu^2-1)(1+ \cos \theta) \sin^2 \gamma \right)
\nonumber \\ 
& & \hspace{2cm} {} - \tilde{g} \frac{2 \pi l_c^2}{a^2} \sqrt{1 - \sin^2 \gamma 
~ \frac{1+\cos \theta}{2}}   \Bigg) 
\end{eqnarray}
$\tilde{g}$ is the Zeeman parameter which is ~60K for $B\sim 45 T$ and $\cos
\theta \equiv \hat{n}_1 \cdot \hat{n}_2$

$E_V$ is minimized at $\gamma = 0$ or $\gamma = \pi$. Thus the nearest
neighbour interaction picks out the $Z_2$ broken, $U(1)$ unbroken spin and
charge ordered state (CDW) discussed previously \cite{alicea}\cite{herbut}. It
is interesting to note that when the Dirac sea contributions are neglected
($\mu=1$), $E_V$ is independent of the $Q$. This was also noticed by Alicea
and Fisher \cite{alicea}. They found that making the interactions slightly
non-local picks out the $\gamma=0,\pi$ state.

The Hubbard term, $E_U$, does not contribute to the mean field energy if we
negelect the contributions from the sea of filled  Dirac-Landau levels. This
can be qualitatively understood in the $n=0$ subspace. Since the two valley
species live on the the two distinct sub-lattices, if only one $n=0$ level is
occupied, we can have any $SU(4)$ polarization without any double occupancies.
However, when the $n\ne 0$ levels are taken into account, $E_U$ is minimized
when $\theta=0$ and $\gamma =\pm \pi/2$. The ground state is thus
anti-ferromagnetic spin ordering (SDW) with the $U(1)$ valley symmetry broken.

We need to minimise this energy density Eq.(\ref{etot}) with respect to the
variational parameters $\theta$, $\gamma$ and $m$. Let us first consider the
case when we ignore the Zeeman energy. The energy density minimizes for
$\gamma =0,\pi$ (independent of $\theta$) and $\gamma=\pm \pi/2$ ( $\theta=0$)
. The minimum energy is obtained at, 

\begin{eqnarray}
\gamma=0(\pi) & \textrm{if} & 3V-U > 0  ~ ; ~ \textrm{any} \; \theta\\
\gamma=\frac{\pi}{2} (-\frac{\pi}{2}) & \textrm{if} & 3V-U < 0  ~;~ \theta=0
\end{eqnarray}

\begin{figure}[ht] \centering
\includegraphics[width=0.45\textwidth]{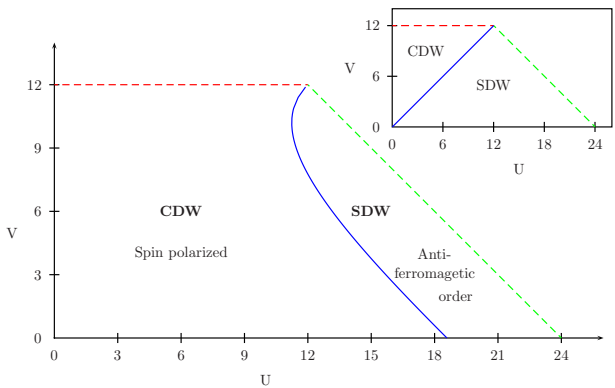}
\caption{The figure shows the phase diagram for $\nu=\pm 1$ in the presence
of the Zeeman term. 
The inset shows the phase diagram for the case when the Zeeman
term is neglected. The CDW region, $Z_2$ part of the valley symmetry is broken.
In the SDW region, $U(1)$ part of the valley symmetry is broken. All energies
are in eV.}
\label{phase-diagram}
\end{figure}

Thus at large $V$ we have the $Z_2$ broken, $U(1)$ unbroken, spin polarised
state and at large $U$ we have the $Z_2$ unbroken, $U(1)$ broken,
anti-ferromagnetic phase.  The transition between them is first order and
occurs at $3V=U$(at mean field level). The phase diagram is shown in inset of
Fig.(\ref{phase-diagram})

The minimization of energy density Eq.(\ref{etot}), yields same two phases in
the $V$-$U$ space discussed above.  The masses in the two phases are given by

\begin{eqnarray}
m_{CDW}&=& \frac{a^2}{2\pi l_c^2} \frac{3V}{4}
\left(1-\frac{3V}{2\tilde{t}}\right)^{-1} \\
m_{SDW}&=& \frac{a^2}{2\pi l_c^2} \frac{(3V+U)}{8}
\left(1-\frac{3V+U}{4\tilde{t}}\right)^{-1} 
\end{eqnarray}

The transition remains first order and the line separating two phases is given
by the solution of 

\begin{eqnarray}
\frac{\left(\frac{3V+U}{4\tilde{t}}\right)^2}{1-\frac{3V+U}{4\tilde{t}}}  
- \frac{\left(\frac{3V}{2\tilde{t}}\right)^2}{1-\frac{3V}{4\tilde{t}}}  
& = & \frac{2\tilde{g}}{\tilde{t}}\frac{2\pi l_c^2}{a^2} 
\end{eqnarray}
Note that to leading order in $a/l_c$, the phase boundary is independent
of the magnetic field. In general we expect a weak magnetic field dependence.
 
The resulting phase diagram for the $\nu=\pm1$ is shown in
Fig.(\ref{phase-diagram}). Thus at large $U$ the Dirac sea polarization
effects can drive the system into a $U(1)$ valley symmetry broken
anti-ferromagnetic phase.  Since the total spin polarization of this state is
zero, the gap will have no dependence on the parallel component of the
magnetic field in this phase, which is consistent with the recent experiments
\cite{jiang}.  We therefore have a possible mechanism to explain the tilted
field experiments at filling factors $\nu=\pm1$ \cite{jiang}.  It has been
argued earlier, the Hubbard $U$ drives the system towards anti-ferromagnetic
order \cite{herbut2}.  It can be seen from the Fig(\ref{phase-diagram}) that
the estimate for the critical Hubbard $U(\sim 15~eV)$ is large. This number
could change by including higher order corrections in $a/l_c$ and the
fluctuations about the mean field. Correlations may also be important even at
the integer fillings \cite{khveshchenko} \cite{baskaran}.

As we mentioned earlier, the dominant interaction in graphene is the $SU(4)$
symmetric long range part of the Coulomb repulsion. As shown by Yang et. al.
\cite{kun-yang}, this interaction favours $SU(4)$ ``ferromagnetic" order by
the direct exchange mechanism. The $SU(4)$ ferromagnetic order is quantified
by the value of $\langle\Psi^\dagger_{\eta\sigma}
\Psi_{\eta^\prime\sigma^\prime}\rangle$. As can be seen from equation
(\ref{gamman0}), this is non-zero for the variational states we are
considering.  It is in fact independent of $m$ because it gets contributions
only from the $n=0$ levels. Thus the dominant gap in graphene will come from
the the Coulomb exchange interaction which is proportional to
$\sqrt{B_\perp}$.

The variational states also have an $SU(4)$ ``anti-ferromagnetic" order
corresponding to a staggering at the scale of the lattice spacing. This is
quantified by the value of $\langle\bar\Psi_{\eta\sigma}
\Psi_{\eta^\prime\sigma^\prime}\rangle$. Equation (\ref{gammann}) shows that
this is the component that is enhanced by a non-zero value of $m$. However at
the scale of $l_c$ over which the exchange mechanism operates, the effects of
this order which varies at the scale of $a$ will be small. The dominant gap
will therefore be the ``ferromagnetic" gap discussed above which is
proportional to $\sqrt{B_\perp}$.

Thus we have shown that if among the short-range part of the interactions, the
Hubbard interaction is dominant, then graphene has a $U(1)$ valley symmetry
broken phase with zero spin polarization at $\nu=1$. The gap in this phase is
proportional to $\sqrt{B_\perp}$. This phase is therefore consistent with the
experimental results reported \cite{jiang} where it is seen that the gap
varies as $\sqrt{B_\perp}$ and is independent of $B_\parallel$.   

There will be three gapless collective modes in this phase. Two corresponding
to the $SU(2)\rightarrow U(1)$ spin symmetry breaking and one corresponding to
the $U(1)$ valley symmetry breaking. The topological defects in this phase are
clearly interesting objects to study.

We have also analysed the $\nu=0$ phase in detail. The state is now
parameterised by two orthogonal $SU(4)$ spinors with ten parameters
corresponding to the $U(4)/(U(2)\times U(2))$ ground state manifold. The
results will be reported in a longer forthcoming publication along with more
details of the results reported in this letter.

\begin{flushleft}
{\bf Acknowledgement:} We thank G. Baskaran for many useful discussions.
\end{flushleft}

\bibliographystyle{apsrev}
\bibliography{ref}
\end{document}